\documentclass[prl,aps,twocolumn,superscriptaddress]{revtex4-1}

\usepackage{graphicx}
\usepackage{bm}
\usepackage{amsfonts}
\usepackage{color}

\newcommand{\St}{\mbox{\textit{St}}} \newcommand{\Steff}{\bar{S}}
\newcommand{\Fr}{\mbox{\textit{Fr}}} \newcommand{\bu}{\boldsymbol{u}}
\newcommand{\bv}{\boldsymbol{v}}
\newcommand{\bvp}{\boldsymbol{v}_\perp}

\newcommand{\xp}{\boldsymbol{X}_\mathrm{p}}
\newcommand{\vp}{\boldsymbol{V}_\mathrm{p}}
\newcommand{\de}{\mathrm{d}}
\newcommand{\dd}[2]{\frac{\de{#1}}{\de{#2}}}

\newcommand{\ocaaddress}{Laboratoire Lagrange, Universit\'e de
  Nice-Sophia Antipolis, CNRS, OCA, Bd.\ de l'Observatoire, 06300
  Nice, France}

\begin{document}

\title{Gravity-driven enhancement of heavy particle clustering in
  turbulent flow}

\author{J\'er\'emie Bec} \affiliation{\ocaaddress}
\author{Holger Homann} \affiliation{\ocaaddress}
\author{Samriddhi Sankar Ray} \affiliation{International Centre for
  Theoretical Sciences, Tata Institute of Fundamental Research,
  Bangalore 560012, India}

\begin{abstract}
  Heavy particles suspended in a turbulent flow settle faster than in
  a still fluid. This effect stems from a preferential sampling of the
  regions where the fluid flows downward and is quantified here as a
  function of the level of turbulence, of particle inertia, and of the
  ratio between gravity and turbulent accelerations. By using
  analytical methods and detailed, state-of-the-art numerical
  simulations, settling is shown to induce an effective horizontal
  two-dimensional dynamics that increases clustering and reduce
  relative velocities between particles. These two competing effects
  can either increase or decrease the geometrical collision rates
  between same-size particles and are crucial for realistic modeling
  of coalescing particles.

\end{abstract}

\maketitle

\noindent Many industrial, atmospheric, and astrophysical phenomena
ranging from the microphysics of cloud formation, to planet formation
in a dusty circumstellar disk of gas, involves the modeling of the
interactions between small solid particles suspended in a turbulent
carrier flow.  Two main effects are typically at play: a viscous drag
that particles experience with the agitated fluid and an external
force, such as gravity, that acts because of their density contrast
with the fluid.  While drag is predominant for small particles,
gravity takes over the dynamics of large particles and most studies
treat these two asymptotics independently.  However it is usually at
this critical transition that standard modeling fails, as is evident
when estimating for instance the rate at which rain is triggered in
warm clouds \cite{FFS02,GW13}. Most models are unable to circumvent a
bottleneck in the droplet growth for diameters around 20-40$\mu{m}$.
A key improvement might be to combine turbulent and gravitational
effects.

In this Letter we understand the intriguing interplay between
turbulence, gravity, and particle sizes. This question is of
fundamental importance in fluid dynamics, in particular, and in
non-equilibrium statistical physics, in general, as it is central to
modeling coalescences in natural or laboratory droplet
suspensions. The most noticeable effect of turbulence on the settling
of heavy particles is the increase of their terminal velocity induced
by a preferential sweeping along the downward fluid flow
\cite{M87,WM93,ayala2008effects}.  This phenomenon is mostly
understood on qualitative grounds and has been quantified only in
model flows \cite{davila2001settling}.  Furthermore very little is
known on the effect of gravitational settling on two-particle
statistics.  Fundamental theoretical and numerical studies of the
clustering of particle pairs~\cite{BFF01,BBC+07} and of the
enhancement of collisions due to inertia \cite{WMB06,BBC+10} usually
neglect gravity.  We present here, by combining state-of-the-art
direct numerical simulations with theoretical results based on our
asymptotic analysis, a systematic study of the dynamical and
statistical properties of particles as a function of \textit{(i)}~the
level of turbulence of the carrier flow (Reynolds number),
\textit{(ii)}~the inertia of the particles (Stokes number), and
\textit{(iii)}~the ratio between the turbulent accelerations and
gravity (Froude number).

We consider a fluid flow whose velocity $\bm u$ is a solution to the
incompressible Navier--Stokes equation
\begin{equation}
  \partial_t \bu + (\bu\cdot\nabla)\bu = -\nabla p + \nu\nabla^2
  \bu +\boldsymbol{f}, \quad \nabla\cdot\bu = 0,
  \label{eq:navier-stokes}
\end{equation}
where $\nu$ is the fluid kinematic viscosity.  Homogeneous isotropic
turbulence is maintained in a statistical steady state by the
large-scale forcing $\boldsymbol{f}$. We perform direct numerical
simulations by using the parallel solver \emph{LaTu}, which is
pseudo-spectral in space and involves a third-order Runge--Kutta
scheme for time marching.  We use different spatial resolutions,
$256^3$, $1024^3$, and $2048^3$ corresponding to Taylor-scale Reynolds
numbers $R_\lambda \!=\! u_\mathrm{rms}\,\sqrt{15/(\varepsilon\nu)}
\approx 130$, $290$, and $460$, respectively ($u_\mathrm{rms} \!=\!
\langle u_x^2\rangle$ is the root-mean-square velocity and
$\varepsilon \!=\!  \nu \langle \|\nabla \bm u \|^2\rangle$ the
average dissipation rate; see~\cite{frisch1996turbulence}).

Particles are assumed much smaller than any turbulent scale, much
heavier than the surrounding fluid, and with a small Reynolds number
associated to their slip velocity. They are then moved by the fluid
through a viscous Stokes drag and their trajectories $\xp(t)$ follow
\begin{equation}
  \dd{\xp}{t} = \vp,\quad \dd{\vp}{t} =
  -\frac{1}{\tau_\mathrm{p}}\left[ \vp-\bu(\xp,t)\right] +
  \boldsymbol{g},
  \label{eq:particles}
\end{equation}
where $\bm g$ is the acceleration of gravity. The relaxation time
reads $\tau_\mathrm{p} \!=\!
2\rho_\mathrm{p}a^2/(9\rho_\mathrm{f}\nu)$, $\rho_\mathrm{p}$ and
$\rho_\mathrm{f}$ being the particle and fluid mass density and $a$
the particle radius. In our simulations individual particle
trajectories are integrated for different values of $\tau_\mathrm{p}$
and $g \!=\! |\bm g|$. The fluid velocity at the location of the
particles is evaluated by linear interpolation. Particle inertia is
measured in terms of the \emph{Stokes number} $\St \!=\!
\tau_\mathrm{p}/\tau_\eta$ where $\tau_\eta \!=\!
\sqrt{\nu/\varepsilon}$ is the turnover time associated to the
Kolmogorov dissipative scale
$\eta\!=\!\nu^{3/4}/\varepsilon^{1/4}$. The effect of gravity is
measured by the \emph{Froude number} $\Fr \!=\!  \varepsilon^{3/4}
/(g\nu^{1/4})$ defined as the ratio between the typical turbulent
acceleration $\eta /\tau_\eta^2$ and that of gravity.  In all our
simulations we have used 10 different Stokes numbers and 5 different
values of the Froude number (including the case of no
gravity). Furthermore, to obtain reliable statistics, we use a very
large number of particles $N_p$, in each of our simulations; for
$R_\lambda \approx 460, N_p = 10^9$, for $R_\lambda \approx 290, N_p =
1.28 \times 10^8$, and $R_\lambda \approx 130, N_p = 8 \times 10^6$.

\begin{figure}
  \begin{center}
    \includegraphics[width=.45\textwidth]{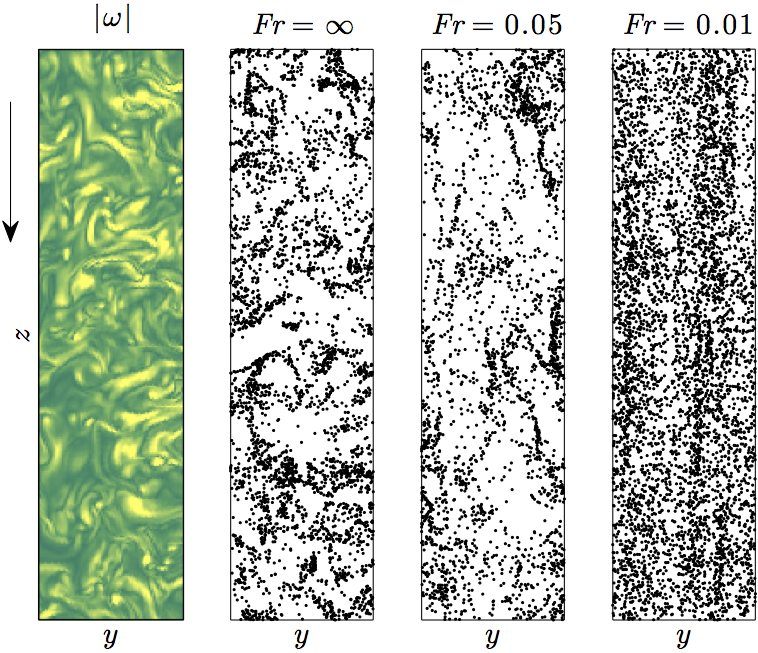}
    \vspace{-10pt}
    \caption{\label{fig:snap1} (color online) Snapshot of the
      vorticity modulus (Left; yellow = low values, green = high
      values) and of the particle positions for $R_\lambda = 130$,
      $\St=1$ and three different values of the Froude number in a
      slice of thickness $10\eta$, width $130\eta$, and height
      $520\eta$. The vertical arrow indicates gravity.}
  \end{center}
\end{figure}
Figure~\ref{fig:snap1} shows a representative snapshot of the modulus
of the vorticity $\boldsymbol{\omega} = \nabla\times\bm u$ in a thin
slice of our three-dimensional flow, together with the position of
particles with the same Stokes ($\St\!=\!1$) but different Froude
numbers $\Fr$. We observe that when gravity is negligible
($\Fr\!=\!\infty$), the particle distribution correlates with
low-vorticity regions. Increasing the effect of gravity does not
instantaneously destroy particle clusters but rather give them some
anisotropy as they get more and more aligned with the vertical
direction $\boldsymbol{\hat{e}}_z \! = \!  -\boldsymbol{g}/g$. This
indicates that settling is responsible for a two-dimensionalization of
the particles dynamics and, as observed on the right-most panel,
clustering is still present but concentration gradients are mainly in
the horizontal directions.

To understand further the underlying mechanisms, let us estimate the
average settling velocity $V_g = -\langle
\vp\cdot\boldsymbol{\hat{e}}_z\rangle$. Clearly from
(\ref{eq:particles}), the statistical stationarity of the particle
velocity implies that $V_g = \tau_\mathrm{p}g - \langle
u_z(\xp,t)\rangle$. The first term is equal to the terminal velocity
of a particle with response time $\tau_\mathrm{p}$ in a still
fluid. It was observed that settling particles in a turbulent flow are
more likely to sample regions where the vertical fluid velocity is
aligned with gravity, leading to an enhancement of their average
settling speed~\cite{M87,WM93}.
\begin{figure}
   \centerline{
      \includegraphics[width=.45\textwidth]{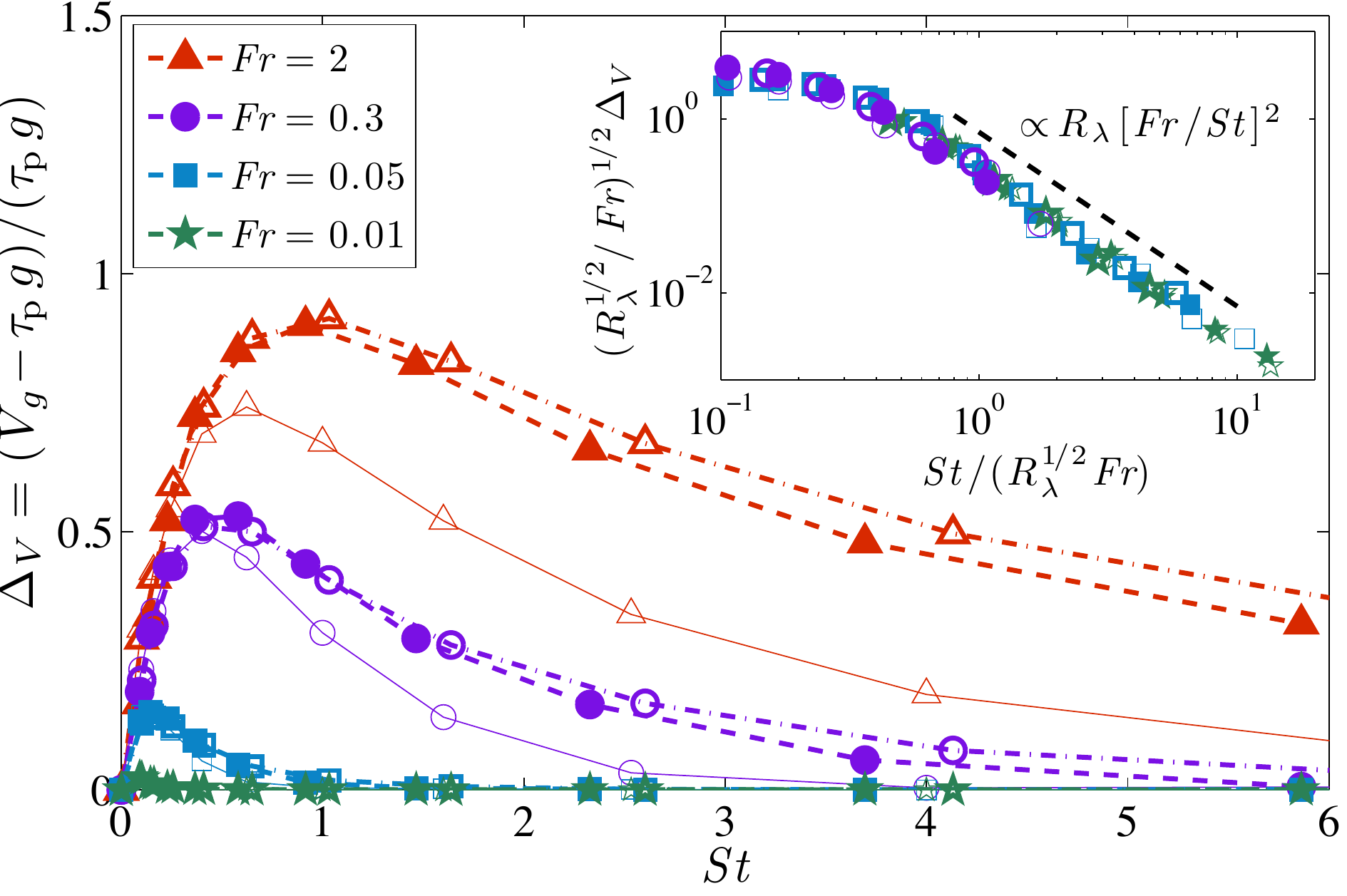}
    }
    \vspace{-10pt}
    \caption{\label{fig:settlingvelocity} (color online) Relative
      increase of the settling velocity $\Delta_{V}$ as a function of
      the Stokes number $\St$ for various Froude numbers, as labeled,
      and $R_\lambda = 130$ (thin symbols, plain lines), $R_\lambda =
      290$ (filled symbols, dashed lines) and $R_\lambda = 460$ (open
      symbols, broken lines). Inset:
      $[R_\lambda^{1/2}/\Fr]^{1/2}\Delta_{V}$ as a function of
      $\St/[R_\lambda^{1/2}\Fr]$ for the same data.}
\end{figure}
This effect is also observed in our simulations as shown in
Fig.~\ref{fig:settlingvelocity} which represents the relative increase
in settling velocity $\Delta_V = (V_g-\tau_\mathrm{p}g) /
(\tau_\mathrm{p}g) = - \langle u_z(\xp,t)\rangle/ (\tau_\mathrm{p}g)$
(compared to the terminal velocity in still fluid) as a function of
the particle Stokes number and for different values of $\Fr$ and
$R_\lambda$. One notices that the effect is the strongest for the
largest values of the Froude number (when turbulent accelerations
dominate over gravity) and, at sufficiently large particle
inertia. Conversely, for low-inertia particles ($\St\ll 1$), the gain
$\Delta_V$ is almost independent of both the Reynolds and the Froude
numbers.

To understand quantitatively these observations, let us first consider
the asymptotics $\St\ll1$. To leading order, the particles are as if
advected by an effective compressible velocity field~\cite{M87},
namely $\vp \approx \boldsymbol{v}(\xp,t)$ with
\begin{equation}
  \bv =  \bu - \tau_\mathrm{p} \left[\partial_t \bu
    +(\bu+\tau_\mathrm{p}\,\boldsymbol{g})\cdot \nabla\bu\right].
  \label{eq:defv}
\end{equation}
We focus on the motion in the horizontal directions $(x,y)$ transverse
to that of gravity and use (\ref{eq:defv}) to write the correlation
$\left\langle u_z\nabla_\perp\cdot\bvp \right\rangle$, where $\bvp
\!=\! (v_x,v_y)$.  All terms except the advection due to settling
vanish by incompressibility or isotropy of the fluid velocity field,
so that
\begin{equation}
  \left\langle u_z\nabla_\perp\cdot \bvp \right\rangle =
  \tau_\mathrm{p}^2 g \left\langle (\partial_z u_z)^2 \right\rangle
  >0. \label{eq:correl}
\end{equation}
Hence, the horizontal clustering of particles (negative divergence) is
on average where the flow heads downward ($u_z\!<\!0$).  This
quantifies the preferential sweeping ideas of~\cite{WM93}. With
arguments similar to those used in~\cite{FFS02} (see
also~\cite{fouxon2012distribution}) for density correlations, we can
relate the average vertical velocity along particle paths to the
correlation (\ref{eq:correl}).  In the limit $\St\ll 1$, the
divergence $\nabla_\perp \cdot \bvp$ gives indeed the bias due to the
preferential sampling by inertial particles.  This leads to $\langle
u_z(\xp,t) \rangle\propto \tau_\eta \langle u_z \nabla_\perp\cdot\bvp
\rangle$ and by using (\ref{eq:correl}) to $\Delta_V \propto
\tau_\eta\tau_\mathrm{p} \left\langle (\partial_z u_z)^2 \right\rangle
\propto \St$, confirming the linear behavior independent of $\Fr$ and
$R_\lambda$ observed in Fig.~\ref{fig:settlingvelocity} at small
Stokes numbers.

In the other asymptotics ($\St\gg1$), the settling velocity $V_g$ gets
very large and the fluid velocity seen by the particles becomes
short-correlated in time. The particles have an almost ballistic
motion in the vertical direction and diffuse in the horizontal
plane. This occurs when the time $L/V_g$, required by the particle to
traverse the integral scale $L = u_\mathrm{rms}^3/\varepsilon$, is much
shorter than the large-scale correlation time $\tau_L =
L/u_\mathrm{rms}$. Rescaling time by $\tau_L(V_g/u_\mathrm{rms})$ and
space by $L$ leads to approximating the horizontal dynamics in
non-dimensional units as
\begin{equation}
  \frac{\mathrm{d}\vp^\perp}{\mathrm{d}s} \simeq
  -\frac{1}{\Steff}\left[\vp^\perp -
    \boldsymbol{\tilde{u}}(\xp^\perp,s)\right]
  \label{eq:effective_onepoint}
\end{equation}
where $\Steff \!=\! (\tau_\mathrm{p}/\tau_L) (u_\mathrm{rms}/V_g)$ and
$\boldsymbol{\tilde{u}}$ is a two-dimensional white-noise-in-time
velocity field whose correlations have the same spatial structure as
$\bm u$.  This approach is similar to that developed
in~\cite{fouxon2008separation} for particles with very large inertia.
Hence, the rescaled single-particle statistical properties depend only
on the effective Stokes number $\Steff$. In particular, the 
falling speed of the particle takes the form $V_g \!\simeq\!  \tau_\mathrm{p}\,g +
u_\mathrm{rms}^2\Psi(\Steff)/V_g$ where $\Psi$ is a non-dimensional
function that accounts for preferential sampling. To leading order
$V_g\simeq\tau_\mathrm{p}\,g$, so that $\Delta_V \sim \Psi(\Steff) \,
R_\lambda \left({\Fr}/{\St}\right)^{2}$ and
$\Steff\sim\Fr/\sqrt{R_\lambda}\ll 1$, which is independent of the
particle Stokes number.  As the velocity field appearing
in~(\ref{eq:effective_onepoint}) is a white-noise, one expects that
for $\Steff\ll1$ statistical observables can be written as a series of
half-integer powers of $\Steff$ (see, \textit{e.g.},
\cite{bec2008stochastic}), so that $\Psi(\Steff) \sim
\Steff^{1/2}$. This leads to
\begin{equation}
  \Delta_V \propto R_\lambda^{3/4} \Fr^{5/3} \St^{-2}
\end{equation}
for $\St\gg R_\lambda^{1/2} \Fr$ and $\Fr\ll R_\lambda^{1/2}$.  This
behavior is confirmed by our data, as can be seen in the inset of
Fig.~\ref{fig:settlingvelocity}.

We now turn to small-scale two-particle statistics.
The approach in terms of the effective dynamics
(\ref{eq:effective_onepoint}) extends to the linearized dynamics,
\textit{i.e.}\ to the tangent system associated to
(\ref{eq:particles}), which describes the evolution of infinitesimal
separations between particles in terms of the fluid gradient
$\nabla\bm u$ along their paths. When the fall speed is large enough,
the particles travel through the correlation length $\eta$ of
$\nabla\bm u$ in a time shorter than its correlation time
$\tau_\eta$. This occurs when $V_g\gg u_ \eta = \eta/\tau_\eta =
(\nu\varepsilon)^{1/4}$. Rescaling time by $\tau_\eta (V_g/u_\eta)$
and space by the Kolmogorov scale $\eta$ allows one to approximate the
time-evolution of the separation $\bm R$ as
\begin{equation}
  \frac{\de^2\boldsymbol R}{\de s^2} \simeq
  -\frac{1}{\tilde{S}}\left[ \dd{\boldsymbol{R}}{s} -
    \boldsymbol{R}\cdot\boldsymbol{\sigma}(s)\right]\!,
  \label{eq:R_lrgeVg}
\end{equation}
where $\boldsymbol{\sigma}$ is a Gaussian tensorial noise with
co-variance $\langle\sigma_{ij}(s)\sigma_{k\ell}(s')\rangle =
(\nu/\varepsilon)\langle \partial_i u_j \partial_k u_\ell \rangle
\delta(s-s')$. The one-point one-time strain tensor co-variance is
here evaluated along particle paths to account for preferential
sampling. The effective Stokes number now reads $\tilde{S} =
\St\,(u_\eta/V_g)$ and in the asymptotics $V_g\gg u_\eta$, small-scale
two-particle statistics depend solely on $\tilde{S}$.  Remarkably,
when the settling velocity is close to that in still fluid
(\textit{i.e.}\ $\Delta_V\ll1$), the effective Stokes reads $\tilde{S}
\simeq \Fr$, so that statistics become independent of $\St$ when
$\St\gg\Fr$.

\begin{figure}
  \begin{center}
    \includegraphics[width=.45\textwidth]{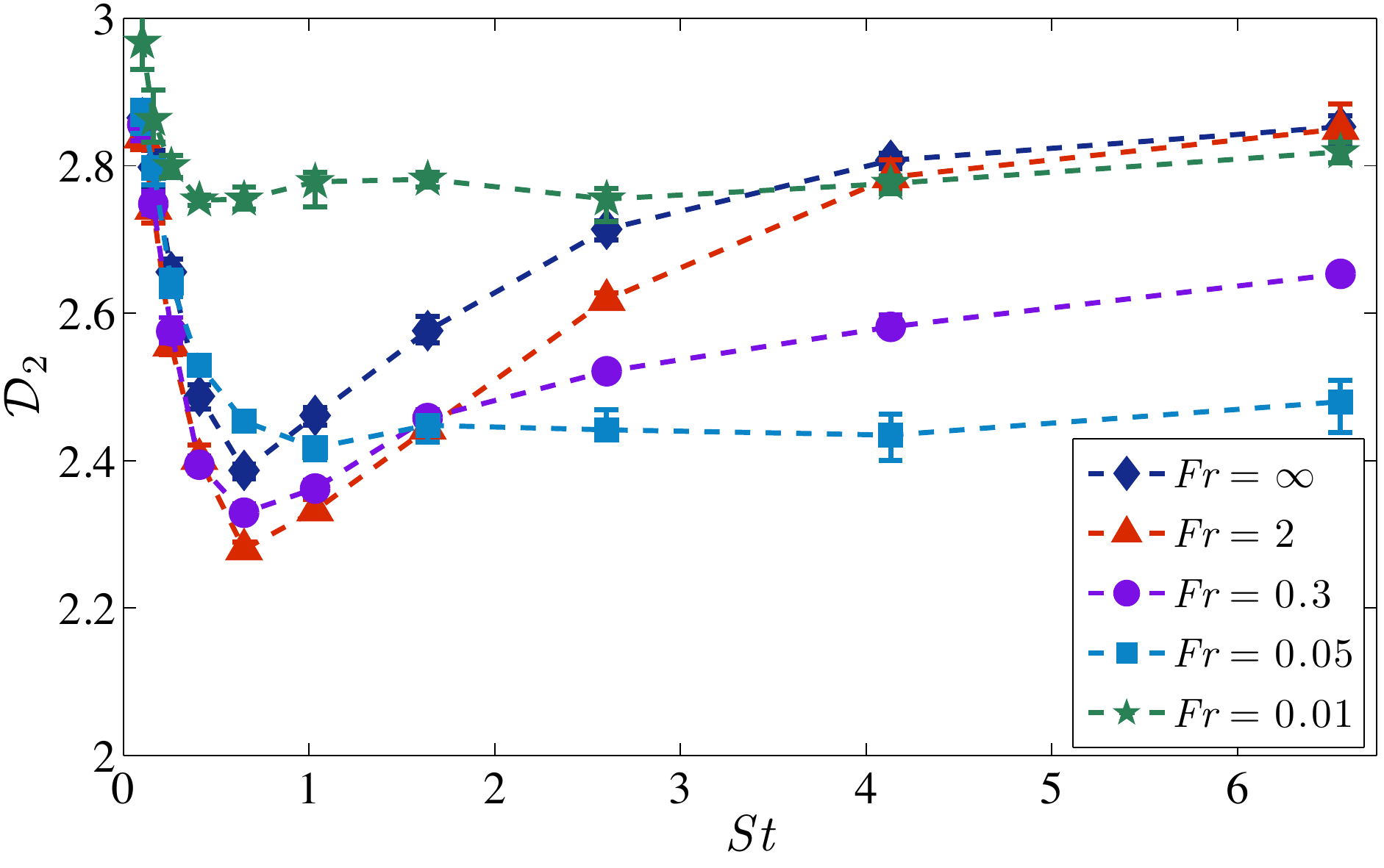}
    \vspace{-10pt}
    \caption{\label{fig:d2} (color online) Correlation dimension
      $\mathcal{D}_2$ of the particle distribution as a function of
      the Stokes number for $R_\lambda = 460$ and various Froude
      numbers as labeled. Smaller Reynolds numbers (not shown here)
      display a similar behavior.}
  \end{center}
\end{figure}
An important observable measuring particle clustering is the
correlation dimension $\mathcal{D}_2$ of their spatial
distribution~\cite{BBC+07}.  It is given by $\mathbb{P}_2(r)\propto
r^{\mathcal{D}_2}$ for $r\ll\eta$, where $\mathbb{P}_2(r)$ is the
probability that two particles are within a distance~$r$.  This
fractal dimension is shown in Fig.~\ref{fig:d2} as a function of $\St$
for different values of $\Fr$ and $R_\lambda = 460$.  One observes
that gravity acts in a non-uniform manner. It tends to enhance
concentration (decrease $\mathcal{D}_2$) when both the Stokes and the
Froude numbers have moderate values. When $\Fr\ll 1$, clustering is
decreased for $\St\lesssim 1$ and increased for $\St\gtrsim1$. For all
finite $\Fr$, one observes that $\mathcal{D}_2$ saturates to a finite
value when $\St\to\infty$. This can be explained by the equivalence
between the two-points dynamics and (\ref{eq:R_lrgeVg}). For $V_g\gg
u_\eta$, the fractal dimension $\mathcal{D}_2$ is a function of the
effective Stokes number $\tilde{S}$ only, which for $\St\gg\Fr$
becomes independent of $\St$.  In this asymptotics, the correlation
dimension depends solely on $\Fr$.  The limiting value of
$\mathcal{D}_2$ is a non-monotonic function of $\Fr$. It is close to 3
when either $\Fr\gg1$ or $\Fr\ll1$ as the corresponding values of
$\Steff\sim\Fr$ characterizing the dynamics (\ref{eq:R_lrgeVg}) are
related in both cases to space-filling distributions.  In
delta-correlated flows, the correlation dimension is known to behave
linearly at small Stokes numbers~\cite{bec2008stochastic}. It is hence
expected that for $\Fr\ll1$, $\mathcal{D}_2\simeq 3-C\,\Fr$ where $C$
is a positive constant.

The increase in clustering observed for order-unity values of $\St$
and $\Fr$ means that settling can significantly impact the timescales
of interaction between particles.  When interested for instance in the
collisions, estimations of the geometrical rate involve the
probability density that two particles are at a distance $r=2a$ equal
to the sum of their radii and thus scales as
$(2a)^{\mathcal{D}_2-1}$. However, this quantity alone is not enough
as the collision rate involves also the typical velocity at which
particles approach each other. Indeed, for same-size particles, it is
given by setting $r=2a$ in the approaching
rate~\cite{bec2005clustering}
\begin{equation}
  \kappa(r) = -\left\langle w\,\theta(-w)\,\delta(|\bm R|-r)\right\rangle,
\end{equation}
where $w \!=\! {\mathrm{d}|\bm R|}/{\mathrm{d}t}$ is the longitudinal
velocity difference between particles, $\theta$ the Heaviside
function, and $\langle\cdot\rangle$ the average over all particle
separations $\bm R$.  The approaching rate can be expressed in terms
of a conditional average $\kappa(r) \!=\! -\left\langle
  w\,\theta(-w)\middle|\,|\bm R|\!=\!r\right\rangle
(\mathrm{d}\mathbb{P}_2/\mathrm{d}r)$, which involves the average
negative longitudinal velocity difference between particles separated
by a distance $r$. This last quantity behaves also as a power of $r$
for $r\ll\eta$ with an exponent $\xi_1$ given by the first-order
structure function of particle velocities~\cite{BBC+10}. This implies
that $\kappa(r)\sim r^\gamma$ with $\gamma =
\xi_1+\mathcal{D}_2-1$. The dependence of $\gamma$ upon $\St$, which
encompasses particle clustering and velocity differences statistics,
determines how the collision rate depends on the particles size and
inertia.
\begin{figure}
  \begin{center}
    \includegraphics[width=.45\textwidth]{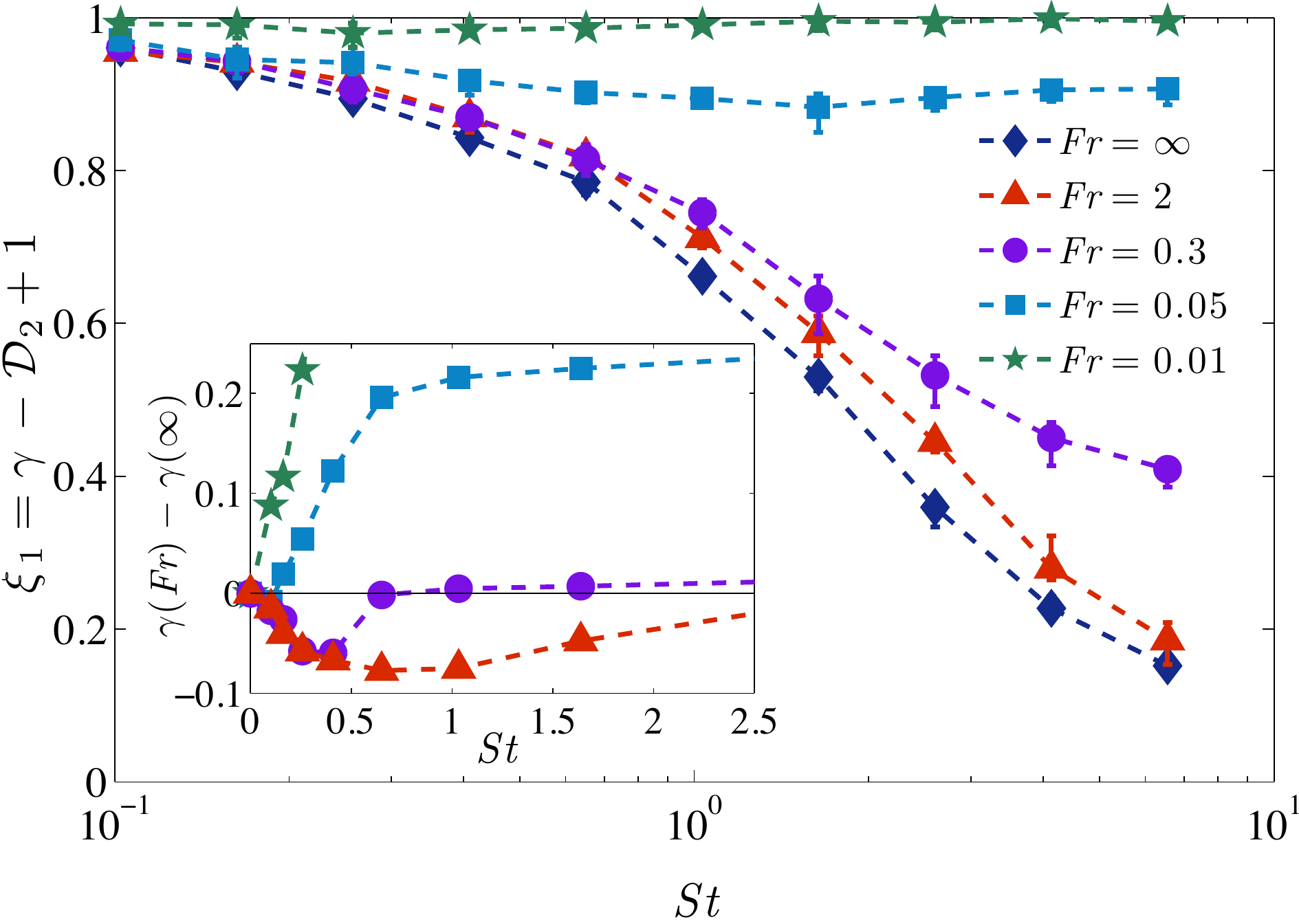}
    \vspace{-10pt}
    \caption{\label{fig:gamma} (color online) Exponent of the velocity
      difference $\xi_1= \gamma-\mathcal{D}_2+1$ as a function of the
      Stokes number for different $\Fr$ and $R_\lambda = 460$. Inset:
      difference between the approaching rate exponent $\gamma$
      associated to the different values of $\Fr$ and that associated
      to particles feeling no gravity ($\Fr=\infty$).  Smaller
      Reynolds numbers (not shown here) behave similarly.}
  \end{center}
\end{figure}
Figure~\ref{fig:gamma} shows the velocity scaling exponent $\xi_1$ as
a function of $\St$ for the various relative strengths of gravity we
have investigated. In the case of no gravity ($\Fr=\infty$), the
particle velocity scaling exponent goes from a behavior close to that
of tracers ($\xi_1=1$) at small $\St$ to an uncorrelated gas with
scale-independent velocity differences ($\xi_1=0$) for
$\St\to\infty$. This transition relates to the formation of fold
caustics in the particle velocity
field~\cite{WMB06,bec2005clustering}. Gravity acts this time in a
monotonic manner since $\xi_1$ systematically increases when $\Fr$
decreases, indicating that settling weakens small-scale velocity
differences between particles.  The underlying mechanisms can be
understood in the asymptotics $\St\gg\Fr$, again in terms of the
equivalent small-scale dynamics (\ref{eq:R_lrgeVg}). When $\Fr$
decreases, the effective Stokes number decreases, so that particles
get closer to tracers of the effective flow and $\xi_1\to 1$.

The two mechanisms determining the rate at which particles collide,
namely preferential concentration and large velocity differences, are
thus affected in competing manners by gravity.  However the
enhancement of particle clustering takes over the decrease of velocity
differences when $\St\lesssim\Fr$. This is evident in the inset of
Fig.~\ref{fig:gamma}, which shows the difference of the approaching
rate scaling exponents $\gamma(\Fr)-\gamma(\infty)$ between particles
undergoing gravity and those that do not. One clearly observes that
$\gamma(\Fr)<\gamma(\infty)$ for $\St\lesssim\Fr$, indicating that in
this range collision rates between same-size particles are larger in
the presence of gravity. At first glance these corrections could seem
tiny. However, they are responsible for an important increase of the
geometrical collision rate. For instance, in highly-turbulent cloud
settings, namely $\Fr=0.3$ (corresponding to $\varepsilon\approx1000\,
cm^2/s$ for the turbulent airflow), we find an increase by more than a
factor two of the collision rate between $\St=0.4$ particles (droplets
with diameter $\approx 30\,\mu m$). This newly identified effect,
which combines turbulence and gravitational settling, clearly needs to
be borne in mind when improving existing models for coalescing
particle or droplet suspensions.

This research has received funding from the European Research Council
under the European Community's Seventh Framework Program
(FP7/2007-2013 Grant Agreement No.~240579) and from the French Agence
Nationale de la Recherche (grant BLAN07-1\_192604). Access to the IBM
BlueGene/P computer JUGENE at the FZ J\"ulich was made available
through the PRACE project PRA031. JB and SSR acknowledge support from
the Indo-French Centre for Applied Mathematics and SSR from the EADS
Corporate Foundation Chair, awarded to ICTS-TIFR and TIFR-CAM.

\bibliography{bib}

\end{document}